Effect of thermal conductivity on the simultaneous formation of a stable region at the top of Earth's core and magnetic field generation over four billion years

Takashi Nakagawa[1,2]*, Shin-ichi Takehiro[3], Youhei Sasaki[4]

[1]Emerging Media Initiative, Kanazawa University, Kanazawa 920-1192, Japan

[2]Department of Earth and Planetary System Science, Hiroshima University, Higashi-Hiroshima 739-8536, Japan

[3]Research Institute for Mathematical Sciences, Kyoto University, Kyoto 606-8502, Japan

[4]Department of Information Media, Hokkaido Information University, Ebetsu 069-8585, Japan

*corresponding author, email: takashi@staff.kanazawa-u.ac.jp

Keywords: Kinetic energy production rate; Long-term magnetic field generation; Stably stratified region; Core-mantle boundary

5633 words with 4 Figures and 2 Tables

**Abstract (213 words).**   The possibility of the emergence of a stratified region in the uppermost part of the Earth's outer core with long-term magnetic field generation is assessed, taking into account uncertainties in the thermal conductivity of the Earth's core and the present-day heat flow across the core-mantle boundary (CMB). The radial structures of the Earth's outer core are calculated for various values of thermal conductivity and CMB heat flow using a one-dimensional thermo-chemical model. The results show that there exist solutions that allow both emergence of stable stratification and long-term magnetic field generation although their thickness of stratified region is thinner than 100 km. In order to satisfy both emergence of stratified region and long-term magnetic field generation, possible value of the present-day CMB heat flow (13-15 TW) suggests a thermal conductivity of 77-121 W/m/K at CMB, which is in good agreement with the values estimated from the electrical conductivity measurements under the Earth's core condition. The thickness of the stratified region in this case is about 50 km, which is also consistent with the thickness of the stratified region estimated from the geomagnetic secular variation. However, the proposed values of thermal conductivity obtained by this analysis could be smaller when the present-day CMB heat flow becomes smaller than the constraint used in this study.

1. Introduction

The existence of a stably stratified region at the top of the Earth's outer core has been suggested to explain the slower seismic wave speed compared to the reference model (e.g., Kaneshima and Matsuzawa, 2015) and 60-year frequency of the geomagnetic secular variation (e.g., Buffett, 2014). However, whether such a region would exist at the top of the Earth's core or not remains controversial, as there are alternative interpretations of seismic wave speed (e.g., Irving et al., 2018) and secular variation (e.g., Gastine et al., 2020) without assuming a stable region at the top of the Earth's core.

The stably stratified region is defined as a region where the radial density gradient is less than the adiabatic one. In order to deal with the formation of a stably stratified layer, some studies have introduced a thermal and/or chemical diffusion layer to one-dimensional models beneath the CMB when the above condition is satisfied (e.g., Labrosse et al., 1997; Lister and Buffett, 1998). Their results suggest that a stable layer can grow to a thickness of 70-140 km when chemical stratification is dominant (Lister and Buffett, 1998; Buffett and Seagle, 2010; Nakagawa, 2018), whereas it can grow to a thickness of about 400 km when thermal stratification is dominant (Labrosse et al., 1997; Greenwood et al., 2021).

Another approach is to examine whether the radial structures calculated by assuming the whole convective core are contradicted or not. Labrosse (2015) calculated the thermal history of the Earth's core using the high thermal conductivity of iron alloys and demonstrated that the radial one-dimensional structure exhibits a region of negative heat transport in the upper part of the outer core, suggesting the potential for a thermal stratification of approximately 1000 km in the upper part of the outer core. Takehiro and Sasaki (2018) proposed the assessment scheme for the stable layer formation using the radial profile of kinetic energy production rate instead of convective heat transport. The advantage is that it can take into account the mixing effect of compositional convection, which was not considered by Labrosse (2015). They suggested that the heat flow across the core-mantle boundary (CMB) should be less than 9.3 TW for emergence of the stratified region at the uppermost outer core. Nakagawa et al. (2023) improved the model of Takehiro and Sasaki (2018) by incorporating downward chemical flux modelled by the core-mantle chemical coupling and the radial profiles of reference density, isentropic temperature and thermal conductivity of the Earth's core. In addition, the strength of magnetic dipole moment was calculated to account for the constraint of continuous magnetic field generation over 4 billion years. The results suggest that stable stratification and magnetic field generation are simultaneously feasible within the range of 11 TW to 14 TW at the present-day CMB heat flow. The maximum thickness of the stratified region in these cases ranges from 50 to 75 km.

These studies (Labrosse, 2015; Takehiro and Sasaki, 2018; Nakagawa et al., 2023) used the radial profile of thermal conductivity of the Earth's core proposed by Gomi et al. (2013), where the value of that is fixed to 163 W/m/K at the centre of the Earth, which corresponds to 89.7 W/m/K at the CMB. However, there is considerable uncertainty about the thermal conductivity of the Earth's core, which ranges from 20 W/m/K to 220 W/m/K at the centre of the Earth's core (Ohta et al., 2016, Konôpková et al., 2016, Hsieh et al., 2020), corresponding to 10-120 W/m/K at the CMB condition based on the radial dependence taken from Labrosse (2015). Furthermore, the constraints on the present-day CMB heat flow have recently been pointed out, with the value of thermal conductivity at the CMB between 75 and 130 W/m/K at the CMB based on the theoretical estimate on the thermal evolution of Earth's core (Frost et al., 2022).

In this study, we extend the work by Nakagawa et al. (2023) and seek the conditions for stable stratification at the uppermost part of the outer core and continuous generation of the Earth's magnetic field, by taking into account uncertainties in the thermal conductivity of iron alloys at high pressure and CMB heat flow. We investigate whether these conditions can be met under the current constraints on CMB heat flow and thermal conductivity of Earth's outer core. Model formulation and analysis procedure are introduced in section 2. Section 3 shows calculated results where both stable region and continuous generation of the magnetic field can be found, and the influence of the uncertainty in the thermal conductivity and CMB heat flow. Section 4 summarizes the results of this study, and further uncertainties in the CMB heat flow and in the existence of a stable region inferred from geophysical observations are discussed.

# 2. Model

## 2.1 Model concept

We follow the same one-dimensional model as that of Nakagawa et al. (2023) is adopted, which reconciles the coexistence of stable region at the outermost outer core and continuous magnetic field generation over 4 billion years. The important model concept is the assessment scheme of the stable layer formation by use of kinetic energy production. Positive and negative kinetic energy production imply the occurrence and suppression of convective motion, respectively. Their radial profile is calculated from those of the thermal and compositional convective fluxes, assuming the emergence of whole layer convection, i.e. the thermal and compositional structures are isentropic and uniform composition, respectively. If there is a region of negative kinetic energy production, this is inconsistent with a convective structure, i.e. the possibility of stable layer formation. More precisely, two types of extreme situations associated with stable layer formation were proposed by Takehiro and Sasaki (2018). One is a minimum penetration case, where the stable layer is determined by the negative region of kinetic energy production. The other is maximum penetration case, where the integral of the kinetic energy in the radial direction is used to determine the maximum position reached by the convective motion (See Fig. 1 in Nakagawa et al., 2023). Although the actual boundary between the stable and convective regions would be located between these two assessments in terms of kinetic energy production rate, we hereby focus on looking at cases for the minimum penetration because we are interested in finding the maximum possible thickness of the stable region with satisfying the long-term magnetic field generation. Furthermore, a light element is assumed to be supplied from the silicate mantle to outer core by the baro-diffusion based on Fick's law (Gubbins and Davies, 2013). In addition, it is assumed that there is no radioactive heating source in the outer core.

Moreover, the continuous magnetic field generation over 4 billion years should be satisfied because of a constraint suggested by the paleomagnetic observations (e.g., Biggin et al., 2015). In Nakagawa et al. (2023), the scaling law of dipole field strength was computed for assessing the continuous magnetic field generation, which requires a continuous positive value of dipole strength over 4 billion years.

## 2.2 Model Calculation Procedure

Here we describe the outline of the calculation procedure in this study. A detailed description of the model formulation can be found in Appendix A.

First, following Labrosse (2015), the reference state of the model is given to be consistent with the Preliminary Reference Earth Model (Dziewonski and Anderson, 1981). The following form is used for the thermal conductivity profile:

$$k_c(r) = k_0 \left[ 1 - A_k \left( \frac{r}{L_\rho} \right)^2 \right] \qquad (1)$$

where $r$ is radius, $k_0$ is the thermal conductivity at the centre of the Earth, $A_k$ is a constant to fit the profile to that of Gomi et al. (2013), and $L_\rho$ is the density scale height defined in Eq. (A6). Here, $k_0$ is treated as a free parameter to investigate the uncertainties of thermal conductivity of Earth's core, while the value of $A_k$ is fixed for simplicity.

Next, for a given CMB heat flow ($Q_{CMB}(t)$) and inner core size ($c$) at a certain time, the growth rate of the inner core $dc/dt$ at that moment is calculated by the global energy balance (See Eq. (A18)):

$$\frac{dc}{dt} = \frac{Q_{CMB}}{[P_c(b) + P_L(b) + P_G(b)]} \qquad (2)$$

where $P_c(b)$, $P_L(b)$ and $P_G(b)$ are the terms related to the secular cooling, latent heat release and gravitational energy release defined in Eqs. (A15) to (A17), and $b$ is the radius of Earth's core. After determining $dc/dt$, the convective entropy flux $F_S(r)$ and convective compositional flux $F_c(r)$ can be diagnostically calculated by the global energy and mass balance equations (Eqs. (A22) and (A25)) as follows.

$$F_c(r) = S_{ICB} - 4\pi \frac{d\bar{X}}{dt} \int_c^r r'^2 \rho_c(r') dr' \qquad (3)$$

$$F_S(r) = \frac{Q_{conv}(r) - \mu'(r) F_C(r)}{T_c(r)} \qquad (4)$$

where $Q_{conv}(r)$ is the convective heat transport, $\mu'(r)$ is the total chemical potential changes, $S_{ICB}$ is the compositional flux at the ICB, $\rho_c(r)$ is the density structure of outer core, and $d\bar{X}/dt$ is the time variations of outer core composition. As a result, the kinetic energy production rate $w(r)$ is determined as:

$$w(r) = g(r) \left( \frac{\alpha_T T_c(r)}{c_p} F_S(r) - \alpha_c F_C(r) \right), \qquad (5)$$

where $g(r)$ is the gravitational acceleration, $\alpha_T$ and $\alpha_C$ are thermal and chemical expansivities, respectively and $c_p$ is the heat capacity.

The stable region can be assessed by where the sign of $w(r)$ changes, which is given as:

$$w(r) > 0: Convective \qquad (6a)$$

$$w(r) \leq 0: Stably\ stratified \qquad (6b)$$

This assessment scheme assumes the minimum penetration case.

The magnetic evolution is calculated by the backward time integration from the present to the early Earth using the present size of the inner core as the initial condition. The magnetic field generation represents the strength of dipole moment $M$ in the following form taken from Olson and Christensen (2006), defined by the modified scaling relationship obtained from numerical dynamo simulations.

$$M = 4\pi b^3 \frac{1}{b-c} \int_c^b \left(\frac{\rho_c(r)}{2\mu_0}\right)^{\frac{1}{2}} \left(\frac{(r-c)\sigma(r)}{4\pi r^2 \rho_c(r)}\right)^{\frac{1}{3}} dr \qquad (7)$$

where $\mu_0$ is the magnetic permeability and $\sigma(r)$ is given by the kinetic energy production rate defined in (5):

$$\sigma(r) = \begin{cases} w(r); \ w(r) > 0 \\ 0; \ w(r) \leq 0 \end{cases} \qquad (8)$$

Eq. (7) is derived by adopting the radially averaged kinetic energy production limited in the positive region to the convective heat flux of the original scaling equation of Olson and Christensen (2006).

2.3 Experimental Setup

The analysis procedure in this study is similar to that described in Nakagawa et al. (2023), except for the incorporation of a parameter survey on thermal conductivity. The present-day CMB heat flow $Q_{CMB}^P$ is varied from 5 to 20 TW in 0.1 TW increments, while the thermal conductivity at the Earth's centre $k_0$ ranges from 20 to 220 W/m/K in 1 W/m/K increments, which is equivalent to 11 to 121 W/m/K at the CMB. Regarding the range of $Q_{CMB}^P$, we include the uncertainties provided in literatures (Lay et al., 2008; Nimmo, 2015; Frost et al., 2022). The range of thermal conductivity is chosen from the experimental constraints (e.g., Ohta et al., 2016) rather than the recent theoretical estimate (Frost et al., 2022), considering the reliability of their accuracy. Other physical parameters used for the calculations are listed in Table 1. Note that thermodynamic parameters of the light element are assumed to be those for the oxygen. Numerical calculations of the structure of the outer core are performed with 2266 grid points (1 km intervals) in the radial direction.

The analysis procedure is described as follows:

1. First, the thermal and compositional structures of the outer core are calculated with the possible present values of the parameters for each value of CMB heat flow and thermal conductivity. For these present structures of the outer core, the radial profiles of kinetic energy production rate are calculated. The emergence of a stable stratified region is then determined where the negative ($w \leq 0$) production rate is found. The thickness of stratified region ($d_s$) is given as $d_s = b - r_s$ where $r_s$ is the radius where the sign of $w(r)$ is changed.

2. In order to ascertain whether it is feasible for a magnetic field to be generated over a period of more than 4 billion years, backward time integration is performed from each of the present-day structure obtained above back to 4.6 billion years ago. The objective is to identify the parameters that permit the coexistence of continuous magnetic field generation and a stratified region. The evaluation of magnetic field generation Eq. (7) is based on the magnitude of the magnetic dipole moment. The heat flow across the CMB, which is an essential quantity for the core evolution, is set as the same way as in Nakagawa et al. (2023), which is given as

$$Q_{CMB}(t) = Q_{CMB}^{P} \exp(At) \qquad (9)$$

where $A = (\ln(Q_{CMB}(4.6\text{Ga})/Q_{CMB}^{P}))/4.6\text{Ga}$ and $Q_{CMB}^{P}$ is the present-day CMB heat flow. The unit of time is taken as billion years. In this study, we investigate cases with $Q_{CMB}(4.6\text{ Ga})/Q_{CMB}^{P} = 2$. This quantity is generally given by the activity of mantle convection, such as in Nakagawa and Tackley (2010) because the fluctuations of CMB heat flow associated with various complicated physics in mantle convection is significant. It is noted that this factor is only given for a moderate change of the CMB heat flow as a function of time. This does not encompass the extreme change of the CMB heat flow in early Earth in numerical mantle convection simulations (e.g., Nakagawa and Tackley, 2010).

3. Two series are tested here: No chemical coupling at the CMB ($S_{CMB} = 0$) and chemical coupling at the CMB ($S_{CMB} \neq 0$).

It is again noted that, in the following discussions, the assumptions of equations (7) and (8) mean that an estimate of the magnetic field strength is maximum at each time. This is because the stable region with the maximum thickness can provide all the kinetic energy production for magnetic field generation.

## 3. Results

Thickness of a stratified region ($d_s$) at the present-day and minimum value of magnetic dipole moment ($\min{(M(t))}$) taken from the time series data of backtrace computation are presented in Figure 1 as a function of thermal conductivity at the CMB ($k_c(b)$) and the present-day CMB heat flow ($Q_{CMB}^P$).The narrow band between the red dotted line ($\min{(M(t))} = 0$) and the black dotted line ($d_s = 0$) represents the region where the continuous generation of the magnetic field and the stratified structure at the top of the present-day outer core can be achieved simultaneously. This suggests that a certain range of $k_c(b)$ can be found depending on $Q_{CMB}^P$,where both formation of a stratified region and long-term generation of magnetic field in the Earth's core are realized. Remarkably, in this band region on the $Q_{CMB}^P - k_c(b)$ plane, the expected thickness of a stratified region is < 100 km, which weakly depends on the choice of $Q_{CMB}^P$ and $k_c(b)$. In the following, we present a sample analysis of one dimensional radial convective structure, using the recent finding on the present-day CMB heat flow integrated by various geophysical observations (Frost et al., 2022).

Based on the diagrams in Fig. 1, we focus on thermal and compositional structure and thermal conductivity in the present-day outer core in range of the present-day CMB heat flow taken from the recent finding on integrating various geophysical observations, which is between 13 and 15 TW (Frost et al., 2022). Figure 2 shows the kinetic energy production rate $w(r)$ as a function of radial distance of the present-day Earth's core for $Q_{CMB}^P$ =15 TW, which includes the profiles in the whole outer core and their zoom-up to check the position of sign change of $w(r)$. For $S_{CMB} = 0$ with $k_c(b) = 120$ W/m/K, the thermal stratification with approximately 500 km thickness is found. This corresponds to where isentropic heat flow is greater than a given present-day CMB heat flow. Such a thermal stratification can be reduced to 50 km due to the compositional convection associated with the release of light element in the inner core growth (Fig. 2a and b). On the other hand, for $S_{CMB} \neq 0$ with $k_c(b) = 99.7$ W/m/K, the stable region can be merged by the injection of light element in the core-mantle chemical coupling rather than the thermal effects. This injection of light element from the CMB forms 47 km thickness of stratification caused by the compositional effect (Fig. 2c and d).

Examining the cases for $Q_{CMB}^P = $13, 14, and 15 TW with the appropriate range of thermal conductivity for continuous magnetic field generation and formation of a stable region, the expected thickness of stably stratified layer becomes 47 and 59 km. (Table 2). This is within the range of thickness of the stable region at the outermost outer core suggested by the geomagnetic secular variation (Buffett, 2014; Buffett et al., 2016). In those cases, the thermal conductivity ranges from 77 to 121 W/m/K at the CMB condition. This suggests that the higher thermal conductivity such as that proposed by Ohta et al. (2016) would be supported rather than the lower conductivity results (Konôpková et al., 2016; Hsieh et al., 2020), although the lower conductivity cannot be ruled out due to the uncertainty in the $Q_{CMB}^P$. Additionally, we also check the end-member cases with $Q_{CMB}^P = $5 TW and 20 TW. For the case with $Q_{CMB}^P = $5TW, the thickness of the stable region becomes up to 190 km with a preferred range of thermal conductivity between 18 and 47 W/m/K at the CMB condition. In contrast, for the case with $Q_{CMB}^P = $20 TW, no successful pairs on stable region and thermal conductivity with the long-term magnetic field generation.

Figure 3 shows the backtrace of core evolution over 4.6 billion years allowing a case for both stable region and continuous magnetic field generation over 4.6 billion years, which includes the inner core size, dipole field strength for the CMB heat flow with $Q_{CMB}^{P} = 15$ TW. The onset of inner core growth occurs around $t = 0.65$ Ga, which is consistent with the recent paleomagnetic observations (e.g., Zhou et al., 2022). However, the age of inner core is still controversial because of a large uncertainty on the paleomagnetic intensity jump used for the determination of inner core nucleation timing (e.g., Biggin et al., 2015) or temporal variation of dipole moment associated with change of buoyancy source distribution for magnetic field generation. Driscoll (2016) argues that geodynamo may have transitioned from a multipolar to dipolar regime, to a weak-field dynamo, and finally to a strong-field dipole around the period of inner core nucleation. In contrast, Landeau et al. (2017) conduct numerical dynamo simulations where the buoyancy forcing is supplied from the CMB and ICB before and after the nucleation, respectively. They show a smooth temporal variation of the dipole strength at the CMB, possibly because a weak magnetic field before the nucleation can produce a comparable dipole strength at the CMB as after the nucleation due to the proximity of the forcing and the observation site. Since our estimate of dipole strength is based on the numerical dynamo solutions with several types of buoyancy source distributions and cannot distinguish such transition of dynamo regimes, the jump of dipole strength exists in Fig. 3. In addition, the dipole strength is a very small value prior to the inner core generation, but is not zero throughout 4.6 billion years, satisfying the long-term magnetic field generation. If the effect of buoyancy type could be incorporated into our estimate, the dipole strength before the onset of inner core growth would be increased, resulting in a weaker jump. The present-day value of dipole strength obtained in this study is comparable to the observational constraint ($8 \times 10^{22}$ Am$^2$, Valet, 2003).

4. Discussion and Summary

In this study, using a radial one-dimensional thermal and compositional model, we explored the solutions where the stratification of the uppermost outer core and continuous magnetic field over 4.6 billion years can coexist as a function of the current CMB heat flux and the thermal conductivity of the outer core. The profiles of kinetic energy production rate were diagnostically evaluated and were used to identify the stable layer and to estimate the strength of the dipole moment. The findings of this study are described below.

1. For the stable stratification above the outer core, as determined by the kinetic energy production rate and magnetic field production at 4.6 billion years, We found the combination of a specific thermal conductivity and $Q_{CMB}^P$ that gives the solutions satisfying both emergence of a stable region and continuous magnetic field generation, forming a narrow band region in $Q_{CMB}^P - k_c(b)$ parameter space (See Fig. 1). The thickness of a stable region and dipole strength of these solutions weakly depends on the choice of both thermal conductivity and $Q_{CMB}^P$.

2. Refer to Table 2, using the more stringent range of CMB heat flow (13 TW-15 TW), an appropriate range of thermal conductivity both for emergence of a stable layer and long-term magnetic field generation is restricted between 77.1 to 121.2 W/m/K. This supports the high thermal conductivity determined by high-pressure experiments (Ohta et al. 2016; Hasegawa et al., 2024) but there is still debate on the appropriate value of thermal conductivity. The thickness of the stable layer ranges from 47 to 59 km, which is within a range of thickness of the stable layer (20-150 km) estimated from geomagnetic secular variation (Buffett, 2014; Buffett et al., 2016).

Note that the range of CMB heat flow used in this study is not fully constrained, leading to uncertainty in the thermal conductivity of the Earth's core. The re-analysis of the geophysical data by Frost et al. (2022) was based on the horizontally averaged thickness of the thermal boundary layer above the CMB. However, seismic tomography models suggest strong heterogeneous structure near CMB (e.g., Masters et al., 2000). A possible cause of this heterogeneity is that oceanic crust segregated from the subducting slab may accumulate above the CMB due to its higher density than the surrounding material, which reduces the estimate of the current CMB heat flow (e.g., Nakagawa and Tackley, 2010). Furthermore, the presence of basal magma oceans in the early stages of Earth's evolution may act as a buffer against CMB heat flow, leading to a reduction in CMB heat flow (e.g. Laneuville et al., 2018). If the lower CMB heat flow is possible, the thermal conductivity is also low value for satisfying both emergence of stratified region and continuous magnetic field generation (See Fig. 1), which might support the direct measurement of thermal conductivity at the high pressure (Konôpková et al., 2016). For example, if the current CMB heat flow is 10 TW the expected range of thermal conductivity is 50-60 W/m/K.

In contrast, the expected thickness of a stable stratified region, about 50 km at most, would not depend on the choice of thermal conductivity and the current CMB heat flow. While this value is consistent with the analysis of geomagnetic secular variation of 20-150 km (e.g. Buffett, 2014), some seismic wave analyses suggest the maximum thickness of a stable region of about 700 km (e.g. Kaneshima and Matsuzawa, 2015), although the interpretation of such seismic structure and uncertainties of the waveform analysis are still controversial on the thickness of stable region ranging from 0 km to a few hundred km (Irving et al., 2018; van Tent et al., 2020).

Our estimate of the stable region at the outermost part of the outer core is favorable for the analysis of the geomagnetic secular variation. Assuming the existence of a stable region, Buffett (2014) observed the 60 years periodicity in the secular variation of the axial dipole, which is the weakest degree in the power spectrum of secular variation, whereas this periodicity was found from a record of 150 years (gufm1 of Jackson et al., 2000). In contrast, we note that it is difficult to reproduce the small-scale geomagnetic field morphology with numerical dynamos that include a stratified layer (Olson et al., 2017; Gastine et al., 2020). Intense flux patches require concentration of field lines by fluid downwellings (Amit, 2014). In addition, the expansion and intensification of the reversed flux patches on the CMB that associate with the decreasing dipole would be attenuated by presence of a stable region due to suppression of field line dispersion by fluid upwelling (Gubbins, 1987).

We used a one-dimensional model to investigate the formation of a stable layer and magnetic field generation, which should be verified by three-dimensional calculations with MHD dynamo models. Horizontally averaged radial structure of temperature, composition and kinetic energy production rate should be compared with one-dimensional calculations. Since it is difficult to use the realistic physical parameters of the Earth's core for three-dimensional calculations, moderate values of the parameter settings are within reach where the main balance is similar to that expected in the realistic Earth situation, the so-called 'path theory' (Aubert et al., 2017).

It is also interesting to consider the effects of laterally heterogeneous CMB heat flow, which would induce a stable layer with lateral variations. The formation of such a stable region has already been suggested by co-density convection calculations in a rotating spherical shell (Mound et al., 2019; Mound and Davies, 2020) and numerical dynamo simulations (Terra-Nova and Amit, 2024), which imply a horizontal thickness variation of O(100) km (Mound and Davies, 2020). By locally adopting our results of the one-dimensional model with different values of CMB heat flow for different locations, we could expect the lateral variation of a stable layer thickness with similar to that obtained by the three-dimensional model. Although the estimates by the one-dimensional model do not include the effects of the global flow induced by the lateral variation of the CMB heat flow, they would become fundamental knowledge when the three-dimensional thermo-chemically driven MHD dynamo calculations with laterally varying CMB heat flow are performed in the future.

We caution that the conclusions drawn so far are not necessarily rigorous because of the simplification of the intrinsic processes and the use of uncertain parameters in our model. First, the chemical coupling at the core-mantle boundary is modelled by a constant chemical flux from the mantle to the core caused by the baro-diffusion. However, the reactions are known to be temperature-dependent through high-temperature and high-pressure experiments on chemical reactions between the silicate mantle and iron alloy (e.g., Frost et al. 2010). For example, we could incorporate chemical flux depending on the CMB temperature for more realistic modeling. Second, in this study the CMB heat flow is expressed as a single simple function to represent the cooling process due to mantle convection. Of course, the inclusion of a full mantle evolution model would significantly change the temporal evolution of the CMB heat flow, which would substantially alter the thermal and chemical evolution of the outer core. However, the backward time integration scheme similar to that used in this study could not be performed with numerical mantle convection calculations involving more complex inherent processes. Instead, we should explore possible solutions with a variety of CMB heat flow histories to make the results more convincing. Finally, the heat producing elements in the outer core (mainly radioactive potassium) are not considered in this study because Labrosse (2015), which we refer to as the base model, discussed that heat producing elements of $O(100)$ ppm would not significantly affect the results. In spite of the caveats for applications to Earth described above, our model is in advantage of exploring the model parameter dependencies and sensitivities. A wider range of parameter experiments could lead to more convincing conclusions about the dynamics and material properties of the outer core.

Appendix: Model formulation

A1. Reference state and thermal conductivity of the Earth's core

Following Labrosse (2015), density, gravitational acceleration and isentropic temperature of the reference state of the Earth's core are expressed by the polynomial form:

$$\rho_c(r) = \rho_0 \left[ 1 - \left( \frac{r}{L_\rho} \right)^2 - A_\rho \left( \frac{r}{L_\rho} \right)^4 \right] \qquad (A1)$$

$$g(r) = \frac{4}{3} \pi G \rho_0 r \left[ 1 - \frac{3}{5} \left( \frac{r}{L_\rho} \right)^2 - \frac{3}{7} A_\rho \left( \frac{r}{L_\rho} \right)^4 \right] \qquad (A2)$$

$$T_c(r) = T_c(c) \left[ \frac{\rho_c(r)}{\rho_c(c)} \right]^\gamma \qquad (A3)$$

where $\rho_c(r)$, $g(r)$, and $T_c(r)$ are the density, gravity, and temperature, respectively. $\rho_0$ is the density at the centre of the Earth, $G$ is the gravitational constant, $c$ is the inner core radius, $\gamma$ is the Grüneisen parameter. $T_m(c)$ is the temperature at the inner core boundary (ICB) given by the melting temperature of Earth's core:

$$T_c(c) = T_{m0} - K_0 \left( \frac{\partial T_m}{\partial P} \right)_X \left( \frac{c}{L_\rho} \right)^2 + \left( \frac{\partial T_m}{\partial X} \right)_P \frac{X_0}{f_c \left( \frac{b}{L_\rho} \right)} \left( \frac{c}{L_\rho} \right)^3 \qquad (A4)$$

where $T_{mo}$ is the melting temperature at the center of Earth's core, $K$ is the bulk modulus, $(\partial T_m / \partial P)_X$ is the pressure gradient of melting temperature, $(\partial T_m / \partial X)_P$ is the compositional gradient of melting temperature, $X_0$ is the reference concentration of the light element of the entire core. The light element is only partitioned into the liquid outer core. $b$ is the radius of the Earth's core, and $f_c(x)$ is a function defined by the integration of the density structure as follows:

$$f_c(x) = x^3 \left(1 - \frac{3}{5}x^2 - \frac{3}{7}A_\rho x^4\right) \qquad (A5)$$

and $L_\rho$ and $A_\rho$ are the density scale height and density fitting constant, respectively.

$$L_\rho = \sqrt{\frac{3K_0}{2\pi G \rho_0^2}}; \ A_\rho = \frac{5K_0' - 13}{10} \qquad (A6)$$

where $K_0'$ is the radial derivative of the bulk modulus. This polynomial formulation with the appropriate choice of the parameters is consistent with the density profile of Earth's outer core in the Preliminary Reference Earth Model (Dziewonski and Anderson, 1981).

Thermal conductivity of Earth's core is given by the following form:

$$k_c(r) = k_0 \left[1 - A_k \left(\frac{r}{L_\rho}\right)^2\right] \qquad (A7)$$

where $k_0$ is the thermal conductivity at the centre of the Earth and $A_k$ is a fitting constant of quadric expression of radial dependence (Gomi et al., 2013; Labrosse, 2015). Here, $k_0$ is treated as a free parameter, ranging from 20 to 220 W/m/K to investigate the uncertainties of thermal conductivity of Earth's core, while we fix the value of $A_k$ for simplicity, which is originally determined using 163 W/m/K at the centre of Earth's core as in Labrosse (2015).

The radial distributions of the reference density and thermal conductivity are shown in Figure A1.

## A2. Evolution of the core structure

At first, in order to evaluate thermal and compositional evolution of the core, the growth rate of the inner core has to be estimated from the global energy balance for given CMB heat flow. The distributions of temperature and composition are successively calculated using the growth rate of the inner core.

The global energy balance at the CMB is given as:

$$Q_{CMB} = Q_c(b) + Q_L(b) + E_G(b) \qquad (A8)$$

where $Q_{CMB}$ is the CMB heat flow, $Q_C(b)$ is the secular cooling term, $Q_L(b)$ is the latent heat release due to the inner core growth and $E_G(b)$ is the gravitational energy release caused by the light element release due to the inner core growth. Those terms in the right-hand side of Eq. (A8) are given as:

$$Q_c(b) = -4\pi \int_0^b r'^2 \rho_c(r') c_p \frac{dT_c(r')}{dt} dr' \qquad (A9)$$

$$Q_L(b) = 4\pi c^2 \rho_c(c) T_c(c) \Delta S \frac{dc}{dt} \qquad (A10)$$

$$E_G(b) = -4\pi \int_c^b r'^2 \rho_c(r') \mu'_{IC}(r') \frac{dc}{dt} dr' \qquad (A11)$$

where $C_p$ is the heat capacity, $\Delta S$ is the entropy change caused by the inner core growth, $dc/dt$ is the inner core growth, and $\mu'_{IC}(r)$ is the change of chemical potential caused by the light element release given as:

$$\mu'_{IC}(r) = -\frac{2}{3}\pi G \rho_0 \alpha_{cI}(r^2 - c^2)\left(1 - \frac{3}{10}\frac{r^2 + c^2}{L_\rho^2}\right) \qquad (A12)$$

$$\alpha_{cI} = \frac{\Delta_X \rho_I}{\rho_c(c)X_0} \qquad (A13)$$

where $\Delta_X \rho_{IC}$ is the density change at the inner core boundary and $X_0$ is the present-day light element concentration in the outer core. The global energy balance described in Eq. (A8) can be written as:

$$Q_{CMB} = [P_c(b) + P_L(b) + P_G(b)]\frac{dc}{dt} \qquad (A14)$$

Here,

$$P_c(b) = -4\pi \int_0^b r'^2 \rho_c(r')c_p \frac{dT_c(r')}{dc}dr' \qquad (A15)$$

$$P_L(b) = 4\pi c^2 \rho_c(c)T_c(c)\Delta S \qquad (A16)$$

$$P_G(b) = -4\pi \int_c^b r'^2 \rho_c(r')\mu'_{IC}(r')dr' \qquad (A17)$$

Hence, the inner core growth rate is calculated for given CMB heat flow as:

$$\frac{dc}{dt} = \frac{Q_{CMB}}{[P_c(b) + P_L(b) + P_G(b)]} \qquad (A18)$$

After the inner core growth rate is determined, the compositional structure for the next step is calculated as:

$$\frac{d\bar{X}}{dt} = \frac{dX_I}{dt} + \frac{dX_O}{dt} = \frac{S_{ICB} + S_{CMB}}{4\pi \int_c^b r^2 \rho_c(r)dr} \qquad (A19)$$

where $S_{ICB}$ and $S_{CMB}$ are compositional fluxes at ICB and CMB, respectively, which

are given as:

$$S_{ICB} = 4\pi c^2 \rho_c(c) X_0 \frac{dc}{dt} \qquad (A20)$$

$$S_{CMB} = 4\pi b^2 \rho_c(b) D_c \frac{\alpha_{cO}}{\mu_O} g(b) \qquad (A21)$$

where $D_c$ is the diffusion coefficient of light element from silicate mantle to metallic core, $\alpha_{cO}$ is the compositional expansion coefficient at the CMB ( $\alpha_{cO} = \Delta_X \rho_O / \rho_c(b) X_0$); $\Delta_X \rho_O$ is the density change across the CMB) and $\mu_O$ is the chemical potential of the light element at the CMB. The temperature structure is calculated from Eqns. (A3) and (A4) with the value of $c$ updated by the given value of inner core growth rate.

A3. Kinetic energy production rate

We use the kinetic energy production rate to judge where the stable region can be found. It is evaluated from convective entropy and compositional fluxes. First, the compositional convective flux can be calculated by the mass balance in the Earth's outer core as follows:

$$4\pi r^2 \rho_c(r) \frac{d\bar{X}}{dt} + \frac{d}{dr} F_c(r) = S_{ICB}\delta(r-c) + S_{CMB}\delta(b-r) \qquad (A22)$$

where $d\bar{X}/dt$ is time variation of the average concentrations of the light components in the outer core.

Integration of the mass balance equation in the radial direction results in the following

expression for:

$$F_c(r) = S_{ICB} - 4\pi \frac{d\bar{X}}{dt} \int_c^r r'^2 \rho_c(r') dr' \qquad (A23)$$

The convective entropy flux is given as:

$$F_S(r) = \frac{Q_{conv}(r) - \mu'(r) F_C(r)}{T_c(r)} \qquad (A24)$$

where $Q_{conv}(r)$ is the convective heat transport described as the global energy balance:

$$Q_{conv}(r) = Q_C(r) + Q_L(r) + E_G(r) + Q_S(r) \qquad (A25)$$

and $\mu'(r)$ is the chemical potential in the Earth's outer core given as:

$$\mu'(r) = \mu'_{IC}(r) + \mu'_{OC}(r) \qquad A(26)$$

Here, $\mu'_{OC}$ are the chemical potential at CMB:

$$\mu'_{OC}(r) = -\frac{2}{3}\pi G \rho_0 \alpha_{cO}(b^2 - r^2)\left(1 - \frac{3}{10}\frac{b^2 + r^2}{L_\rho^2}\right) \qquad (A27)$$

Each term in the convective heat transport can be evaluated diagnostically as:

$$Q_c(r) = -4\pi \int_0^r r'^2 \rho_c(r') c_p \frac{dT_c(r')}{dt} dr' \qquad (A28)$$

$$Q_L(r) = 4\pi c^2 \rho_c(c) T_c(c) \Delta S \frac{dc}{dt} \qquad (A29)$$

$$E_G(r) = -4\pi \left(\int_c^r r'^2 \rho_c(r') \mu'_{IC}(r') \frac{dc}{dt} dr'\right) \qquad (A30)$$

$$Q_S(r) = 4\pi r^2 k_c(r) \frac{dT_c(r)}{dr} \qquad (A31)$$

It is noted that the thermal boundary condition at ICB is the isentropic.

With convective entropy and compositional fluxes, the work by buoyancy, that is the kinetic energy production rate in convection, can be described as:

$$w(r) = g(r) \left( \frac{\alpha_T T_c(r)}{c_p} F_S(r) - \alpha_c F_C(r) \right) \qquad (A32)$$

where $\alpha_T$ is the thermal expansion and $\alpha_C$ is the chemical expansion. Using Eq. (A32), we judge convective stability as follows:

$$w(r) > 0: Convective \qquad (A33a)$$

$$w(r) \leq 0: Stably\ stratified \qquad (A33b)$$

It is noted that the maximum thickness of stable region (the assumption on the minimum penetration of convective flow in the Earth's core of Takehiro and Sasaki, 2018 and Nakagawa et al., 2023) is addressed here. An actual boundary between convective and stable regions may be in between the positions corresponding to the minimum penetration and to the maximum penetration (Takehiro and Sasaki, 2018).

A4. Magnetic evolution

As in Nakagawa et al. (2023), the magnetic field generation over 4 billion years is a key component for explaining the structure and dynamics of the Earth's core. In order to measure how the magnetic field can be generated by the convective dynamics of the Earth's core, we use the scaling relationship between work by buoyancy and dipole field strength based on the numerical dynamo simulations (Olson and Christensen, 2006), which is given as:

$$M = 4\pi b^3 \frac{1}{b-c} \int_c^b \left(\frac{\rho_c(r)}{2\mu_0}\right)^{\frac{1}{2}} \left(\frac{(r-c)\sigma(r)}{4\pi r^2 \rho_c(r)}\right)^{\frac{1}{3}} dr \qquad (A34)$$

where $\mu_0$ (H/m) is the magnetic permeability. We define $\sigma(r)$ using the buoyancy flux:

$$\sigma(r) = \begin{cases} w(r); \ w(r) > 0 \\ 0; \ w(r) \le 0 \end{cases} \qquad (A35)$$

Note that Eq. (A34) were originally formulated for the whole region in the Earth's core to estimate the dipole field strength. Our Eq. (A34) leads to find the maximum strength of magnetic field generation because it is assumed that all positive $w(r)$ contributes magnetic field generation. Otherwise, a certain portion of positive kinetic energy may be used for penetrative convection, reducing the kinetic energy of core convection for generating the magnetic field.

**Acknowledgements.** We thank Hagay Amit and anonymous reviewer for providing insightful comments. The present study was financially supported by the JSPS Kaken-hi (15H05834, 16H01117, 19H01947, 20K04050, 21H01155, 23K20878, 24K00694). This study was also supported by the Research Institute for Mathematical Sciences, a Joint Usage/Research Center located in Kyoto University.

**Declaration of generative AI and AI-assisted technologies in the writing process.**
During the preparation of this manuscript, the authors used 'DeepL' for improving English usage. After using this tool, the authors reviewed and edited the content as needed and take full responsibility for the content of manuscript.

**Data availability.** Figures shown here can be reproduced by following all the equations and physical parameters given in this study.

*Author contributions (CRediT information)*:

**Takashi Nakagawa:** Conceptualization, Methodology, Software, Formal analyses, Investigation, Writing – Original draft preparation, Writing – Review and editing.

**Shin-ichi Takehiro:** Conceptualization, Methodology, Funding acquisition, Writing – Review and editing

**Youhei Sasaki:** Conceptualization, Methodology, Funding acquisition, Writing – Review and editing

Tables and captions

Table 1. Parameters for the reference structure of Earth's core. The oxygen is assumed as the major light element in this study.

| Notation | Parameter | Value | Reference |
|---|---|---|---|
| $\rho_0$ | Density at the centre | 12451 kg m$^{-3}$ | Labrosse (2015) |
| $L_\rho$ | Density scale height | 8039 km | Labrosse (2015) |
| $A_\rho$ | 4$^{th}$ order polynomial fitting constant of the density | 0.484 | Labrosse (2015) |
| $K_0$ | Bulk modulus | $1.4 \times 10^{12}$ Pa | Labrosse (2015) |
| $K_0'$ | Pressure derivative of Bulk modulus | 3.567 | Labrosse (2015) |
| $b$ | Core radius | 3486 km | Labrosse (2015) |
| $c$ | Inner core radius | 1221 km at the present | Dziewonski and Anderson (1981) |
| $A_k$ | Radial dependence of thermal conductivity | 2.39 | Gomi et al. (2013) |
| $\gamma$ | Grüneisen parameter | 1.5 | Vocaldo et al. (2003) |
| $\left(\dfrac{\partial T_m}{\partial P}\right)_X$ | Pressure derivative of melting temperature | $9 \times 10^{-9}$ K/Pa | Alfe et al. (2007) |
| $\left(\dfrac{\partial T_m}{\partial X}\right)_P$ | Compositional derivative of melting temperature | $-2.1 \times 10^4$ K$^{-1}$ | Alfe et al. (1999) |
| $X_O$ | Reference concentration of light element of the Earth's core | 5.6 % | Labrosse (2015) |
| $\Delta S$ | Entropy change caused by the inner core growth | 127 J/K | Labrosse (2015) |
| $\alpha_{cO}$ | Compositional expansion across the CMB | 1.13 | Hirose et al. (2017) |
| $\alpha_{cI}$ | Compositional expansion across the ICB | 0.83 | Gubbins et al. (2003) |
| $\Delta_X \rho_O$ | Density difference caused by the core-mantle chemical coupling | Computed from $\alpha_{cO}$ | |
| $\Delta_X \rho_I$ | Density difference across the ICB | Computed from $\alpha_{cI}$ | |
| $c_p$ | Heat capacity | 750 J K$^{-1}$ kg$^{-1}$ | Labrosse (2015) |
| $T_{m0}$ | Melting temperature at the centre | 5300 K | Within a range provided by Nomura et al. (2014) and Morard et al. (2014) |
| $\alpha_T$ | Thermal expansion | $10^{-5}$ | Buffett and Seagle (2010) |
| $D_c$ | Chemical diffusivity | $3 \times 10^{-8}$ m$^2$ s$^{-1}$ | Maximum value in Posner et al. (2017) |
| $\mu_O$ | Chemical potential of oxygen | $1.6 \times 10^7$ J/kg | Gubbins and Davies (2013) |

Table 2. Summary of computed results incorporating the recent accomplishment on the CMB heat flow range pointed out by the integrated investigation on various geophysical observations (Frost et al., 2022) plus two end-member cases of CMB heat flow used in this study. N/A means that no successful conductivity which can find both stable region formation and continuous magnetic field generation.

| $Q_{CMB}^P$ (TW) | CMB chemical coupling | Range of thermal conductivity at CMB (W/m/K) | Maximum thickness of stratified region (km) |
|---|---|---|---|
| **Cases taken from integrated geophysical observations** | | | |
| **13** | No | 97.5-105.2 | 59 |
| **13** | Yes | 77.1-85.3 | 58 |
| **14** | No | 105.2-112.3 | 52 |
| **14** | Yes | 84.8-92.5 | 52 |
| **15** | No | 112.3-121.2 | 50 |
| **15** | Yes | 92.5-99.7 | 47 |
| **End-member cases** | | | |
| **5** | No | 17.6-27.0 | 185 |
| **5** | Yes | 37.4-47.3 | 158 |
| **20** | No | N/A | 0 |
| **20** | Yes | N/A | 0 |

Figures and captions

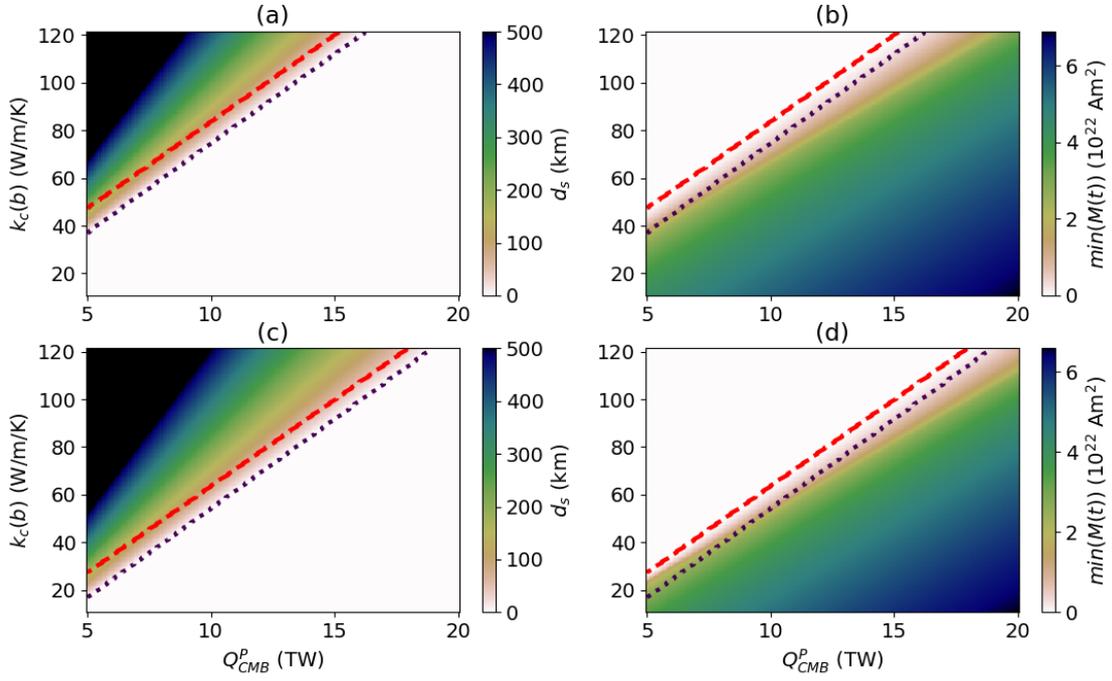

Figure 1. Present-day thickness of a stratified region ($d_s$, left panels) and minimum value of dipole strength taken from the time series data of back trace computations (min ($M(t)$), right panels as functions of $Q_{CMB}^p$ and $k_c(b)$. (a) and (b) indicate the case for $S_{CMB} = 0$. (c) and (d) indicate the case for $S_{CMB} \neq 0$. Black dotted lines indicate the boundary where a stratified layer emerges. Red dashed lines indicate the boundaries where magnetic field generation over 4 billion years is obtained.

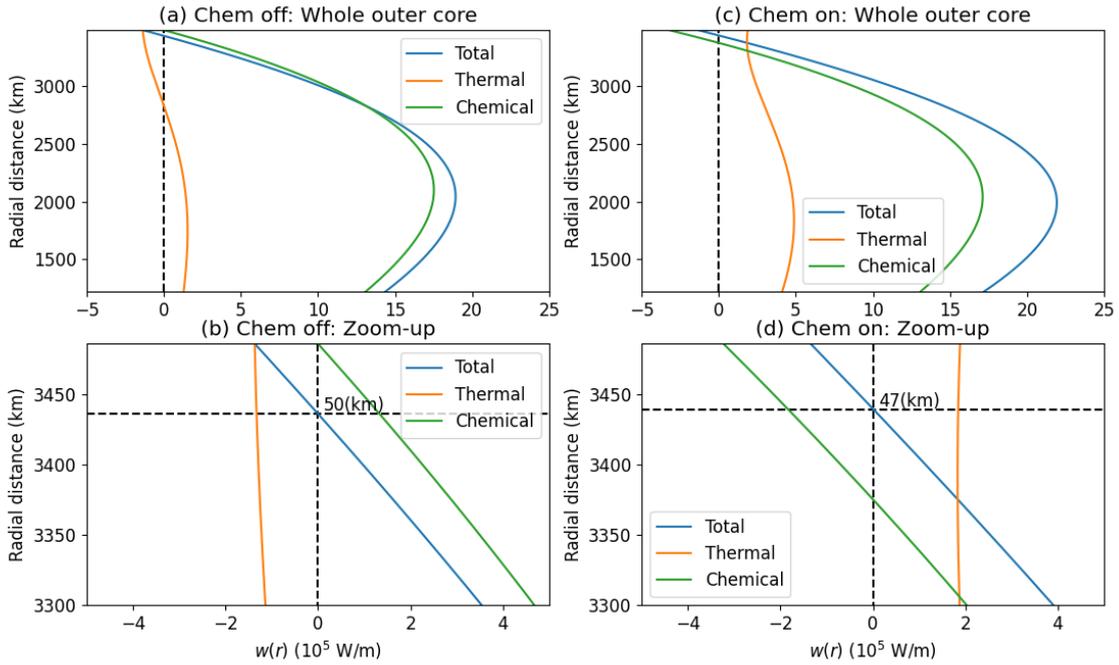

Figure 2. Kinetic energy production rate $w(r)$ as a function of radial distance of the present-day Earth's core for $Q_{CMB}^{P} = 15$ TW. Left and right panels are for $S_{CMB} = 0$ with $k_c(b) = 120.$ and $S_{CMB} \neq 0$ with $k_c(b) = 99.7$, respectively. (a) and (c) show the whole outer core, while (b) and (d) are zoom-in above 3300 km. In the zoom-in figures, the horizonal and vertical dashed lines highlight a stable region.

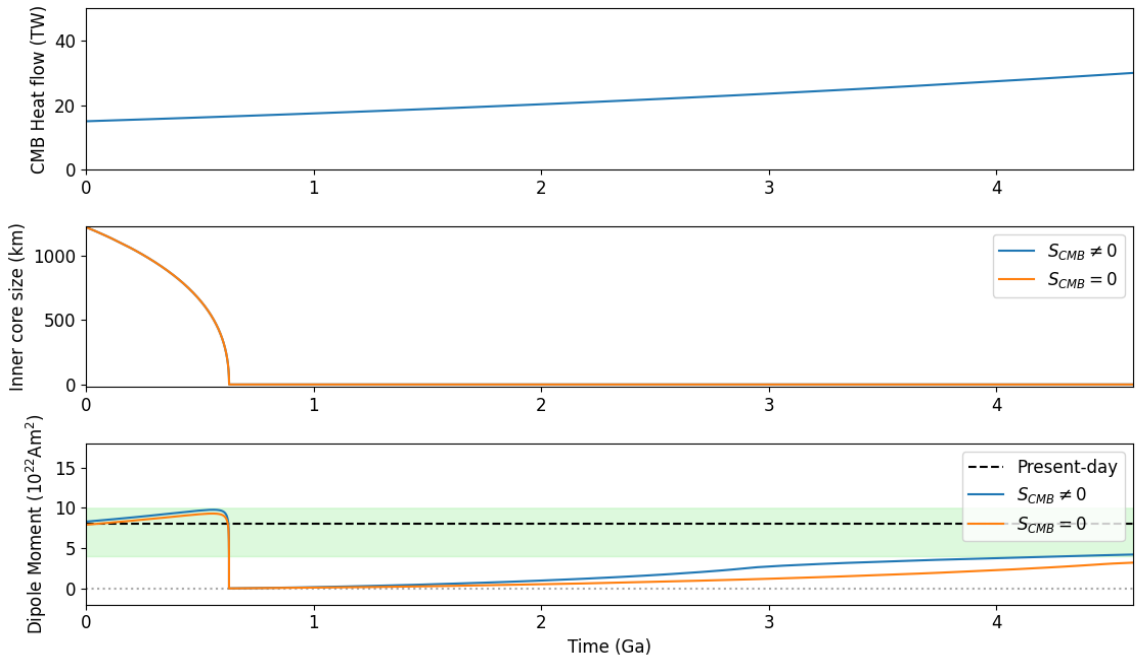

Figure 3. Example of the backward integration in time in the cases of $Q_{CMB}^P = 15$ TW, with $S_{CMB} = 0$ and $k_c(b) = 120.$ and $S_{CMB} \neq 0$ and $k_c(b) = 99.7$, respectively. The dashed line indicates the present-day strength of the dipole moment, $8 \times 10^{22}$ Am$^2$ (e.g., Valet, 2003), and the shaded region indicates the measurement uncertainty of the dipole moment strength (e.g., Biggin et al., 2015). From top to bottom, CMB heat flow, inner core size and strength of magnetic dipole moment are shown.

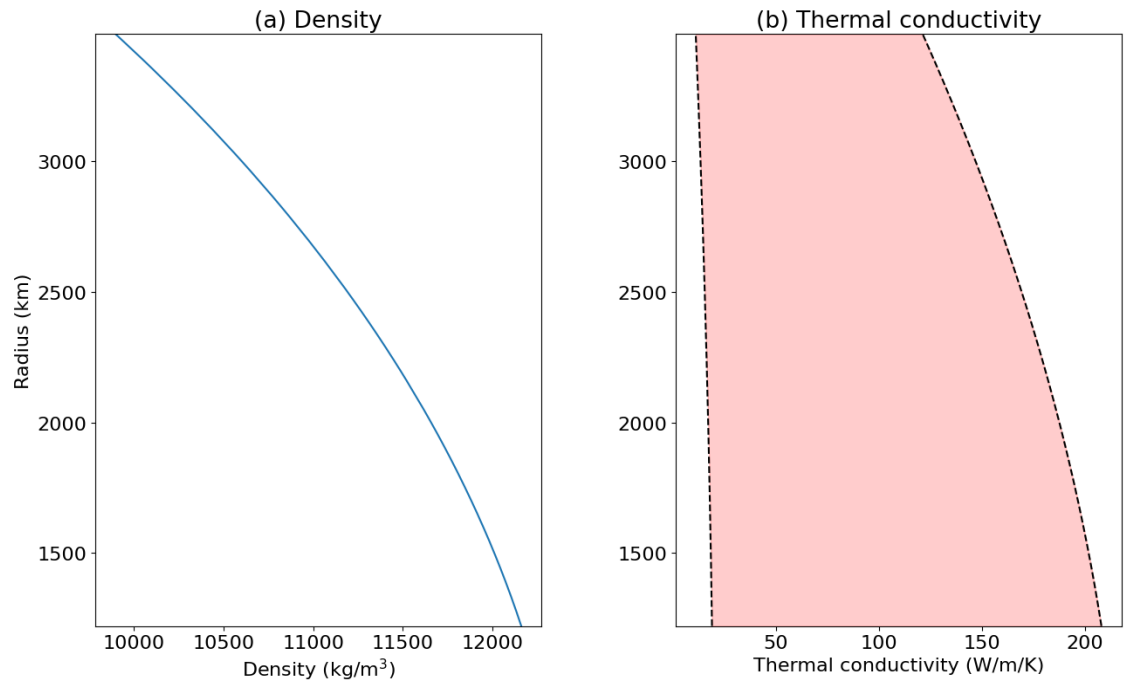

Figure A1. Radial distributions of (a) reference density and (b) thermal conductivity in the outer core. Shaded region in thermal conductivity indicates the uncertainties suggested by high pressure experiments (Ohta et al., 2016; Konopkova et al., 2016; Hsieh et al., 2020).